\documentclass[english,aps,preprint]{revtex4}
\usepackage[T1]{fontenc}
\usepackage[latin9]{inputenc}
\usepackage{babel}
\usepackage{amsmath}
\usepackage{graphicx}
\usepackage{esint}
\usepackage[unicode=true,pdfusetitle,
 bookmarks=true,bookmarksnumbered=false,bookmarksopen=false,
 breaklinks=false,pdfborder={0 0 1},backref=false,colorlinks=false]
 {hyperref}
\usepackage{breakurl}

\makeatletter

\providecommand{\tabularnewline}{\\}

\@ifundefined{textcolor}{}
{%
 \definecolor{BLACK}{gray}{0}
 \definecolor{WHITE}{gray}{1}
 \definecolor{RED}{rgb}{1,0,0}
 \definecolor{GREEN}{rgb}{0,1,0}
 \definecolor{BLUE}{rgb}{0,0,1}
 \definecolor{CYAN}{cmyk}{1,0,0,0}
 \definecolor{MAGENTA}{cmyk}{0,1,0,0}
 \definecolor{YELLOW}{cmyk}{0,0,1,0}
}

\usepackage{simplewick}
\usepackage{babel}

\makeatother

\begin{document}

\title{Nielsen Identity and the Renormalization Group Functions in an Abelian
Supersymmetric Chern-Simons Model in the Superfield Formalism}

\author{A. G. Quinto}
\email{andres.quinto@ufabc.edu.br}

\author{A. F. Ferrari}
\email{alysson.ferrari@ufabc.edu.br}

\affiliation{\emph{Universidade Federal do ABC - UFABC, Rua Santa Adélia, 166,
09210-170, Santo André, SP, Brazil}}
\begin{abstract}
In this paper we study the Nielsen identity for the supersymmetric
Chern-Simons-matter model in the superfield formalism, in three spacetime
dimensions. The Nielsen identity is essential to understand the gauge
invariance of the symmetry breaking mechanism, and it is obtained
by using the BRST invariance of the model. We discuss the technical
difficulties in applying this identity to the complete effective superpotential,
but we show how we can study in detail the gauge independence of one
part of the effective superpotential, $K_{eff}$. We calculate the
renormalization group functions of the model for arbitrary gauge-fixing
parameter, finding them to be independent of the gauge choice. This
result can be used to argue that $K_{eff}$ also does not depend on
the gauge parameter. We discuss the possibility of the extension of
these results to the complete effective superpotential.
\end{abstract}

\maketitle

\section{Introduction}

The effective potential is used to calculate physically meaningful
quantities such as the masses for physical particles, and therefore
its possible gauge dependence is an important question that have been
studied in the literature for quite some time. Given that physical
observables cannot depend on the gauge choice, it is essential to
understand how to extract gauge independent information from perturbative
calculations of the effective action in gauge theories\,\cite{Jackiw:1974cv,Dolan1974,Frere1975}.
A very robust formalism to address this question was developed by
Nielsen, Kudo and Fukuda\,\cite{Nielsen:1975fs,Fukuda:1975di}, providing
identities that encode the behavior of the effective action under
changes of the gauge-fixing parameter. The so-called Nielsen identities
imply that the gauge dependence of the effective action is compensated
by a non-local field redefinition. For the effective potential, for
example, the Nielsen identity reads
\begin{align}
\left(\alpha\,\frac{\partial}{\partial\alpha}+C\left(\sigma;\alpha\right)\frac{\partial}{\partial\sigma}\right)V_{eff}\left(\sigma;\alpha\right) & =0\thinspace,\label{eq:NielsenIdentity}
\end{align}
where $\sigma$ is the vacuum expectation value of the scalar field,
$\alpha$ is the gauge-fixing parameter, $V_{eff}$ the quantum effective
potential, and $C\left(\sigma;\alpha\right)$ is a function which
can be calculated in term of Feynman diagrams. A consequence of this
relation is that physical quantities defined at extrema of the effective
potential become gauge-independent\,\cite{Nielsen:1975fs,Aitchison:1983ns,Johnston1985}.
We find in the literature many examples of the application of the
Nielsen identities in condensed matter physics, QCD, QED, the Standard
Model, ABJM theory, to name a few\,\cite{DoNascimento1987,Breckenridge1995,Gambino2000,Iguri2001,Gerhold2003,Lewandowski2013,Upadhyay2016}. 

In the context of the Chern-Simons (CS) models, there have been reports
of computations performed in a specific gauge, such as the evaluation
of the renormalization group functions presented in\,\cite{Avdeev:1991za,Avdeev:1992jt},
which considered the models both with and without supersymmetry, assuming
as true the gauge invariance. However, one must keep in mind that,
on general grounds, renormalization group functions can depend on
the choice of the gauge-fixing parameter\,\cite{Collins_Book,Fazio2001,Bell2013,DiLuzio2014,Bell2015}.
Many other recent works in the literature also presented calculations
in a specific gauge, without discussing the question of gauge dependence,
such as studies regarding the effective superpotential of supersymmetric
CS models coupled to matter\,\cite{lehum:2007nf,Lehum2009,Ferrari:2009zx,Ferrari:2010ex,Queiruga:2015fzn,Quinto2016},
one exception being\,\cite{Lehum:2010tt}, which considered the large
$N$ limit. Since the effective potential can be dependent on the
gauge-fixing parameter in general\,\cite{Jackiw:1974cv,Dolan1974},
a proper study of gauge independence in CS models is still lacking.

Following theses ideas, our first goal is to study the Nielsen identity
in the context of supersymmetric CS theory, working in the superfield
formalism\,\cite{gates:1983nr,buchbinder:1998qv}, in which the supersymmetry
is manifest in all stages of the calculation. We start by finding
the general Becchi-Rouet-Stora-Tyutin (BRST) transformations associated
to this theory, then use this result to obtain the Nielsen identity
for the effective superpotential $V_{eff}^{S}$. The detailed development
of the Nielsen formalism in the superfield language is the first result
of this work. However, the direct application of this identity for
superfield models in three spacetime dimensions is complicated by
the difficulty in calculating the complete effective superpotential
$V_{eff}^{S}$ in the superfield language\,\cite{Ferrari:2009zx}.
As a first step in this direction, we consider the part of the effective
superpotential which does not depend on supercovariant derivatives
of the background scalar superfield, $K_{eff}$. We calculate the
renormalization group functions with arbitrary gauge-fixing parameter,
and we show explicitly that these are indeed gauge independent. Since
$K_{eff}$ can be calculated from these functions as shown in\,\cite{Quinto2016},
it follows that $K_{eff}$ also does not depend on the gauge choice.
We discuss how to extend these results to the complete effective superpotential
$V_{eff}^{S}$.

This paper is organized as follows: in Section\,\ref{sec:Building-an-invariant-Lagrangian},
we present our model and study its invariance under BRST transformations.
These results are used in Section\,\ref{sec:Calculating-the-Nielsen-Identity}
to find the Nielsen identity in the superfield formalism. The question
of the gauge (in)dependence of the effective superpotential is discussed
in Section\,\eqref{sec:On-the-gauge}. In Section\,\ref{sec:Calculation-of-beta-function},
we calculate the renormalization group functions in the scale invariant
version of our model, up to two loops, with an arbitrary gauge-fixing
parameter. We find these functions to be gauge independent: this,
together with the results of the other sections, allows us to firmly
establish the gauge independence of part of the effective superpotential.
Section\,\ref{sec:Conclusion} presents our conclusions and perspectives.
The two-loops integrals needed for our calculations are presented
in the Appendix.

\section{\label{sec:Building-an-invariant-Lagrangian}Building an invariant
Lagrangian under BRST transformations}

In this section we investigate the BRST transformations in a $\mathcal{N}=1$
supersymmetric Chern-Simons-matter model in $\left(2+1\right)$ dimensions.
Our starting point is the action
\begin{align}
\mathcal{S}_{CS} & =\int d^{5}z\left\{ -\frac{1}{2}\Gamma^{\alpha}W_{\alpha}-\frac{1}{2}\overline{\nabla^{\alpha}\Phi}\nabla_{\alpha}\Phi-m\,\overline{\Phi}\Phi+\frac{\lambda}{4}\left(\overline{\Phi}\Phi\right)^{2}\right\} ,\label{eq:actionCSM}
\end{align}
where $W_{\alpha}=\frac{1}{2}D^{\beta}D_{\alpha}\Gamma_{\beta}$ is
the gauge superfield strength, $\nabla^{\alpha}=\left(D^{\alpha}-ig\,\Gamma^{\alpha}\right)$
is the gauge supercovariant derivative, $D^{\alpha}=\partial^{\alpha}+i\theta_{\beta}\partial^{\beta\alpha}$
is the usual supersymmetric supercovariant derivative and $m$ is
a mass parameter. In this section we will work with a non vanishing
mass parameter $m$ (Higgs model), instead of a theory with conformal
invariance at the classical level (Coleman-Weinberg\,\cite{Coleman:1973jx}
model), both for the sake of generality, and also because the $m=0$
case would be more complicated to analyze in the context of the Nielsen
identities\,\cite{Kang1974,Nielsen:1975fs,Aitchison:1983ns}. 

The action\,\ref{eq:actionCSM} is invariant under the gauge transformations
\begin{eqnarray}
\delta_{g}\Phi=ig\mathcal{K}\Phi,\: & \delta_{g}\overline{\Phi}=-ig\mathcal{K}\overline{\Phi},\: & \delta_{g}\Gamma_{\alpha}=D_{\alpha}\mathcal{K}\thinspace,\label{eq:GaugeTrans}
\end{eqnarray}
where $\mathcal{K}$ is a scalar superfield; these can be rewritten
as a BRST transformation, 
\begin{eqnarray}
\delta_{B}\Phi=i\epsilon gC\Phi,\: & \delta_{B}\overline{\Phi}=-i\epsilon gC\overline{\Phi},\: & \delta_{B}\Gamma_{\alpha}=-\epsilon D_{\alpha}C\thinspace,\label{eq:BRSTTrans-1}
\end{eqnarray}
where $C$ is a ghost superfield and $\epsilon$ is an infinitesimal
constant parameter, both being Grassmannian. As for the BRST transformation
of the ghosts fields, since we consider an Abelian model\,\cite{Becchi1974},
we have
\begin{align}
\delta_{B}C & =0,\thinspace\delta_{B}\overline{C}=\epsilon B\left(z\right),\label{eq:BRSTTrans-2}
\end{align}
$B\left(z\right)$ being a scalar superfield, known as the Lautrup-Nakanishi
auxiliary field in quantum field theory. The BRST transformations
are nilpotent\,\cite{Johnston1985}, i.e.,
\begin{equation}
\delta_{B}^{2}C=\delta_{B}^{2}\overline{C}=\delta_{B}^{2}\Gamma_{\alpha}=\delta_{B}^{2}\Phi=\delta_{B}^{2}\overline{\Phi}=0\,,
\end{equation}
which implies that $\delta_{B}B\left(z\right)=0.$ 

Now let us write the following Lagrangian,
\begin{align}
\mathcal{L}_{t} & =\mathcal{L}_{CSM}+\delta_{B}\mathcal{O},\label{eq:Lagran-t}
\end{align}
which is invariant by the BRST transformations; here, 
\begin{align}
\mathcal{O}\left(z\right) & =\overline{C}\left(z\right)\left(-\alpha\,\frac{1}{4}B\left(z\right)+\frac{1}{2}\,F\left(z\right)\right),
\end{align}
where $F\left(z\right)$ is a gauge fixing function and $\alpha$
the gauge-fixing parameter. Applying the BRST transformation on the
operator $\mathcal{O}\left(z\right)$, we have
\begin{align}
\delta_{B}\mathcal{O} & =-\epsilon\,\frac{\alpha}{4}B^{2}+\epsilon\frac{1}{2}B\,F+\frac{1}{2}\overline{C}\left(\delta F\right)\thinspace.\label{eq:Delta-O}
\end{align}
Integrating out the superfield $B\left(z\right)$, we obtain
\begin{align}
\mathcal{O}\left(z\right) & =\frac{1}{4}\,\overline{C}\left(z\right)\,F\left(z\right),\label{eq:Operador-O}
\end{align}
and
\begin{align}
\delta_{B}\mathcal{O} & =\epsilon\,\frac{1}{4\alpha}\,F^{2}+\frac{1}{2}\,\overline{C}\left(\delta F\right).\label{eq:DeltaBRST-O}
\end{align}
Therefore, the total Lagrangian, after setting $\epsilon=1$, and
redefining the scalar superfield as $\Phi=\frac{1}{\sqrt{2}}\left(\Phi_{1}+i\,\Phi_{2}\right)$,
reads
\begin{align}
\mathcal{L}_{t} & =\frac{1}{2}\,\Gamma_{\alpha}W^{\alpha}+\frac{1}{2}\left(\Phi_{1}D^{2}\Phi_{1}+\Phi_{2}D^{2}\Phi_{2}\right)-\frac{1}{2}m\left(\Phi_{1}^{2}+\Phi_{2}^{2}\right)-\frac{1}{4}g^{2}C^{\alpha\beta}\Gamma_{\beta}\Gamma_{\alpha}\left(\Phi_{1}^{2}+\Phi_{2}^{2}\right)\nonumber \\
 & +\frac{1}{2}\,g\left[\Phi_{1}D^{\alpha}\Phi_{2}-\Phi_{2}D^{\alpha}\Phi_{1}\right]\Gamma_{\alpha}+\frac{\lambda}{16}\,\left(\Phi_{1}^{2}+\Phi_{2}^{2}\right)^{2}+\frac{1}{4\alpha}\,F^{2}+\frac{1}{2}\,\overline{C}\,\frac{\delta F}{\delta\mathcal{K}}\,C\thinspace.\label{eq:Lagran-Total-1}
\end{align}

An useful class of gauge fixing conditions, which is the supersymmetric
generalization of the $R_{\xi}$ gauge, is given by 
\begin{align}
F & =D^{\alpha}\Gamma_{\alpha}+d\,g\,\Phi_{2},\label{eq:Fixing-Gauge}
\end{align}
where $d$ is an arbitrary parameter that can be chosen to eliminate
the mixing between the $\Phi_{2}$ and $\Gamma_{\alpha}$, for example\,\cite{Ferrari:2010ex,lehum:2007nf}.
We leave the value of $d$ unspecified, in which case in general one
would need to consider mixed propagators to evaluate quantum corrections\,\cite{Nielsen:1975fs,Aitchison:1983ns,Johnston1985,DoNascimento1987}.
With this choice of gauge fixing, we end up with the Lagrangian
\begin{align}
\mathcal{L}_{t} & =\frac{1}{4}\Gamma_{\alpha}W^{\alpha}+\frac{1}{2}\left(\Phi_{1}\left[D^{2}-m\right]\Phi_{1}+\Phi_{2}\left[D^{2}-m\right]\Phi_{2}\right)\nonumber \\
 & -\frac{1}{4}\,g^{2}C^{\alpha\beta}\Gamma_{\beta}\Gamma_{\alpha}\left(\Phi_{1}^{2}+\Phi_{2}^{2}\right)+\frac{1}{2}\,g\left[\Phi_{1}D^{\alpha}\Phi_{2}-\Phi_{2}D^{\alpha}\Phi_{1}\right]\Gamma_{\alpha}\nonumber \\
 & +\frac{\lambda}{16}\left(\Phi_{1}^{2}+\Phi_{2}^{2}\right)^{2}+\frac{1}{4\alpha}\,F^{2}+\overline{C}\,D^{2}\,C+\frac{1}{2}\,d\,g^{2}\,\Phi_{1}\overline{C}C,\label{eq:Lagran-Total-2}
\end{align}
which is invariant under the BRST transformations:\begin{subequations}
\begin{align}
\delta_{B}\Gamma_{\alpha} & =-\epsilon D_{\alpha}C,\\
\delta_{B}\Phi_{1} & =-\epsilon g\Phi_{2}C,\\
\delta_{B}\Phi_{2} & =\epsilon g\Phi_{1}C,\\
\delta_{B}\overline{C} & =-\epsilon\frac{1}{\alpha}\,F\,,\\
\delta_{B}C & =0\,.
\end{align}
\end{subequations}Finally, we add to the Lagrangian the source terms
\begin{equation}
\mathcal{L}_{source}=J^{\mu}\Gamma_{\mu}+\overline{\eta}C+\overline{C}\eta+f_{1}\Phi_{1}+f_{2}\Phi_{2}-gK_{1}\Phi_{2}C+gK_{2}\Phi_{1}C+h\mathcal{O}\thinspace,\label{eq:Lagran-Fonte}
\end{equation}
where $J^{\mu},\,\overline{\eta},\,\eta,\,f_{1}$ and $f_{2}$ are
the sources of the basic superfields, while $K_{1},\,K_{2}$ and $h$
are sources for the composite operators.

\section{\label{sec:Calculating-the-Nielsen-Identity}Obtaining the Nielsen
Identity}

Our starting point is the generating functional,
\begin{align}
Z\left[J_{a}\right] & =e^{iW\left[J_{a}\right]}=N\int\mathcal{D}\phi_{a}e^{iS},\label{eq:Func-Geradora-1}
\end{align}
where $J_{a}$ represent all the sources and $\mathcal{D}\phi_{a}$
the path integral over all superfields that are present in the action
\begin{equation}
S=\int d^{5}z\,\left(\mathcal{L}_{t}+\mathcal{L}_{source}\right).
\end{equation}
 Applying the BRST transformations on $W\left[J_{a}\right]$, the
invariance of $\mathcal{L}_{t}$ implies that
\begin{align}
0 & =\frac{1}{Z\left[J_{a}\right]}N\int\mathcal{D}\phi_{a}e^{iS}\int d^{5}z\left(\delta_{B}\mathcal{L}_{source}\right)\,,\label{eq:BRST-Fonte-1}
\end{align}
where
\begin{align}
\delta_{B}\mathcal{L}_{source} & =\epsilon J^{\mu}D_{\mu}C-\epsilon\frac{1}{\alpha}\,F\eta-\epsilon gf_{1}\Phi_{2}C+\epsilon gf_{2}\Phi_{1}C+h\epsilon\tilde{\mathcal{O}}\,,\label{eq:BRST-Fonte-2}
\end{align}
and 
\begin{equation}
\delta_{B}\mathcal{O}=\epsilon\tilde{\mathcal{O}}\thinspace.\label{eq:defOtilde}
\end{equation}
We also quote the useful relations \begin{subequations}\label{eq:VMeio-Superfields}
\begin{eqnarray}
\frac{\delta W\left[J_{a}\right]}{\delta J^{\mu}}=\Gamma_{\mu}\,, &  & \frac{\delta W\left[J_{a}\right]}{\delta\eta}=-\overline{C}\,,\\
\frac{\delta W\left[J_{a}\right]}{\delta\overline{\eta}}=C\,, &  & \frac{\delta W\left[J_{a}\right]}{\delta f_{1}}=\Phi_{1}\,,\\
\frac{\delta W\left[J_{a}\right]}{\delta f_{2}}=\Phi_{2}\,, &  & \frac{\delta W\left[J_{a}\right]}{\delta K_{1}}=-g\,\Phi_{2}C\,,\\
\frac{\delta W\left[J_{a}\right]}{\delta K_{2}}=g\,\Phi_{1}C\,.
\end{eqnarray}
\end{subequations}

The quantum effective action is defined by means of a partial Legendre
transformation,
\begin{align}
\Gamma\left[\phi_{a}\right] & =W\left[J_{a}\right]-\int d^{5}z\left(J^{\mu}\Gamma_{\mu}+\overline{\eta}C+\overline{C}\eta+f_{1}\Phi_{1}+f_{2}\Phi_{2}\right)\,,\label{eq:Funcional-efetiva}
\end{align}
where the sources $h$ and $K_{i}$ are not Legendre transformed.
Besides the usual relations,\begin{subequations}\label{eq:Valor-Fontes}
\begin{eqnarray}
\frac{\delta\Gamma\left[\phi_{a}\right]}{\delta\Gamma_{\mu}}=J^{\mu}\,, &  & \frac{\delta\Gamma\left[\phi_{a}\right]}{\delta\overline{C}}=\eta\,,\\
\frac{\delta\Gamma\left[\phi_{a}\right]}{\delta C}=-\overline{\eta}\,, &  & \frac{\delta\Gamma\left[\phi_{a}\right]}{\delta\Phi_{1}}=-f_{1}\,,\\
\frac{\delta\Gamma\left[\phi_{a}\right]}{\delta\Phi_{2}}=-f_{2}\,,
\end{eqnarray}
\end{subequations}one can also prove that\,\cite{Aitchison:1983ns},\begin{subequations}\label{eq:Relacoes}
\begin{align}
\frac{\delta\Gamma\left[\phi_{a}\right]}{\delta h} & =\frac{\delta W\left[J_{a}\right]}{\delta h}\,,\\
\frac{\delta\Gamma\left[\phi_{a}\right]}{\delta K_{i}} & =\frac{\delta W\left[J_{a}\right]}{\delta K_{i}}\,,\\
\frac{\partial\Gamma\left[\phi_{a}\right]}{\partial\alpha} & =\frac{\partial W\left[J_{a}\right]}{\partial\alpha}\,.
\end{align}
\end{subequations}Using these relations, Eq.\,\eqref{eq:BRST-Fonte-1}
can be cast as
\begin{align}
-\frac{1}{Z\left[J_{a}\right]}N\int\mathcal{D}\phi_{a}e^{iS}\int d^{5}z\left(h\tilde{\mathcal{O}}\right) & =\int d^{5}z\left(\frac{\delta\Gamma\left[\phi_{a}\right]}{\delta\Gamma_{\mu}}D_{\mu}C-\frac{1}{\alpha}\,F\frac{\delta\Gamma\left[\phi_{a}\right]}{\delta\overline{C}}\right.\nonumber \\
 & \left.-\frac{\delta\Gamma\left[\phi_{a}\right]}{\delta\Phi_{1}}\frac{\delta\Gamma\left[\phi_{a}\right]}{\delta K_{1}}-\frac{\delta\Gamma\left[\phi_{a}\right]}{\delta\Phi_{2}}\frac{\delta\Gamma\left[\phi_{a}\right]}{\delta K_{2}}\right)\,,\label{eq:BRST-Fonte-4}
\end{align}
which after functional differentiation with respect to $h$ and taking
$h=0$, reduces to
\begin{align}
-\frac{1}{Z\left[J_{a}\right]}N\int\mathcal{D}\phi_{a}e^{iS}\tilde{\mathcal{O}}= & \int d^{5}z\left(\frac{\delta\Gamma\left[\mathcal{O}\left(z\right)\right]}{\delta\Gamma_{\mu}}D_{\mu}C-\frac{1}{\alpha}\,F\frac{\delta\Gamma\left[\mathcal{O}\left(z\right)\right]}{\delta\overline{C}}\right.\nonumber \\
 & -\frac{\delta\Gamma\left[\mathcal{O}\left(z\right)\right]}{\delta\Phi_{1}}\frac{\delta\Gamma\left[J_{a}\right]}{\delta K_{1}}-\frac{\delta\Gamma\left[\phi_{a}\right]}{\delta\Phi_{1}}\frac{\delta\Gamma\left[\mathcal{O}\left(z\right)\right]}{\delta K_{1}}\nonumber \\
 & \left.-\frac{\delta\Gamma\left[\mathcal{O}\left(z\right)\right]}{\delta\Phi_{2}}\frac{\delta\Gamma\left[\phi_{a}\right]}{\delta K_{2}}-\frac{\delta\Gamma\left[\phi_{a}\right]}{\delta\Phi_{2}}\frac{\delta\Gamma\left[\mathcal{O}\left(z\right)\right]}{\delta K_{2}}\right)\,,\label{eq:BRST-Fonte-5}
\end{align}
where
\begin{align}
\delta\Gamma\left[\mathcal{O}\left(z\right)\right] & =\left.\frac{\delta\Gamma\left[\phi_{a}\right]}{\delta h}\right|_{h=0}\,.\label{eq:BRST-Fonte-5-Complemento}
\end{align}

The operator $\tilde{\mathcal{O}}$ in \eqref{eq:BRST-Fonte-5}, in
the class of supersymmetric $R_{\xi}$ gauges we are considering,
is given explicitly by
\begin{align}
\tilde{\mathcal{O}} & =\frac{1}{4\alpha}\,F^{2}+\overline{C}D^{2}C+\frac{1}{2}d\,g^{2}\,\Phi_{1}\overline{C}C\,,\label{eq:Operador-O-bar}
\end{align}
which, by using the equation of motion $\delta S/\delta\overline{C}=0,$
reduces to
\begin{align}
\tilde{\mathcal{O}} & =\frac{1}{4\alpha}\,F^{2}-\overline{C}\eta\,.\label{eq:Operador-O-bar-1}
\end{align}
By differentiation of $W\left[J\right]=-i\ln Z\left[J\right]$ with
respect to the gauge parameter $\alpha$, considering Eq.\,\eqref{eq:Lagran-Total-2},
one obtains that
\begin{align}
\alpha\frac{\partial W\left[J_{a}\right]}{\partial\alpha} & =\frac{1}{Z\left[J_{a}\right]}N\int\mathcal{D}\phi_{a}\left(-\int d^{5}z\,\frac{1}{4\alpha}\,F^{2}\right)e^{iS}\,,\label{eq:Parte-1-Operador}
\end{align}
and proceeding similarly, 
\begin{align}
-\int d^{5}r\,\eta\left(r\right)\frac{\delta W\left[J_{a}\right]}{\delta\eta\left(r\right)} & =\frac{1}{Z\left[J_{a}\right]}N\int\mathcal{D}\phi_{a}e^{iS}\left(\int d^{5}r\,\eta\left(r\right)\overline{C}\left(r\right)\right)\,.\label{eq:Parte-2-Operador}
\end{align}
These relations, together with Eqs.\,\eqref{eq:Relacoes}, \eqref{eq:Valor-Fontes}
and \eqref{eq:VMeio-Superfields}, allows us to rewrite Eq.\,\eqref{eq:BRST-Fonte-5}
as
\begin{align}
\alpha\frac{\partial\Gamma\left[\phi_{a}\right]}{\partial\alpha}+\int d^{5}z\,\frac{\delta\Gamma\left[\phi_{a}\right]}{\delta\overline{C}}\,\overline{C} & =\int d^{5}r\int d^{5}z\left(\frac{\delta\Gamma\left[\mathcal{O}\left(z\right)\right]}{\delta\Gamma_{\mu}}D_{\mu}C-\frac{1}{\alpha}\,F\frac{\delta\Gamma\left[\mathcal{O}\left(z\right)\right]}{\delta\overline{C}}\right.\nonumber \\
 & -\frac{\delta\Gamma\left[\mathcal{O}\left(z\right)\right]}{\delta\Phi_{1}}\frac{\delta\Gamma\left[\phi_{a}\right]}{\delta K_{1}}-\frac{\delta\Gamma\left[\phi_{a}\right]}{\delta\Phi_{1}}\frac{\delta\Gamma\left[\mathcal{O}\left(z\right)\right]}{\delta K_{1}}-\frac{\delta\Gamma\left[\mathcal{O}\left(z\right)\right]}{\delta\Phi_{2}}\frac{\delta\Gamma\left[\phi_{a}\right]}{\delta K_{2}}\nonumber \\
 & \left.-\frac{\delta\Gamma\left[\phi_{a}\right]}{\delta\Phi_{2}}\frac{\delta\Gamma\left[\mathcal{O}\left(z\right)\right]}{\delta K_{2}}\right)\,.\label{eq:BRST-Fonte-final}
\end{align}
This expression is the base for obtaining the Nielsen identity.

The effective superpotential is obtained by setting $\Phi_{1}=\sigma_{cl}$
in the effective action, 
\begin{equation}
\left.\Gamma\left[\Phi_{1},\alpha\right]\right|_{\Phi_{1}=\sigma_{cl}}=V_{eff}^{S}\left(\sigma_{cl},\,\alpha\right)\thinspace,
\end{equation}
where
\begin{equation}
\sigma_{cl}=\sigma_{1}-\theta^{2}\sigma_{2}
\end{equation}
is the spacetime constant expectation value of the scalar superfield.
By taking $\overline{C}=C=\Phi_{2}=\Gamma_{\alpha}=0$ in Eq.\,\eqref{eq:BRST-Fonte-final},
after some manipulations, we end up with
\begin{align}
\left[\alpha\frac{\partial}{\partial\alpha}+C^{S}\left(\sigma_{cl},\alpha\right)\frac{\partial}{\partial\sigma_{cl}}\right]V_{eff}^{S}\left(\sigma_{cl},\,\alpha\right) & =0,\label{eq:Identidade-Nielsen-V}
\end{align}
where 
\begin{align}
C^{S}\left(\sigma_{cl},\alpha\right) & =\int d^{5}z\left.\frac{\delta^{2}\Gamma\left[\mathcal{O}\left(z\right)\right]}{\delta K_{1}\left(0\right)\delta h\left(y\right)}\right|_{K_{1}=h=0},\label{eq:C-Phi-alpha}
\end{align}
which is the Nielsen identity for the superpotential.

\section{\label{sec:On-the-gauge}On the gauge (in)dependence of the Effective
Superpotential}

The effective superpotential in three spacetime dimensions have the
general form
\begin{align}
V_{eff}^{S}\left(\sigma_{cl},\,\alpha\right) & =-\int d^{5}z\,\left[K_{eff}\left(\sigma_{cl},\,\alpha\right)+\mathcal{F}\left(D_{\alpha}\sigma_{cl},\,D^{\alpha}\sigma_{cl},\,D^{2}\sigma_{cl},\,\sigma_{cl},\alpha\right)\right]\,,\label{eq:SuperPotencialCompleto}
\end{align}
where we made explicit the potential gauge dependence. Here, $K_{eff}$
is the part of the effective superpotential that do not depend on
derivatives of the background classical superfield $\sigma_{cl}$,
similarly to the Kälerian effective superpotential defined in four
dimensional models\,\cite{gates:1983nr,buchbinder:1998qv}. In the
context of dynamical gauge symmetry breaking, it is enough to consider
only the $K_{eff}$\,\cite{Ferrari:2010ex,Quinto2016,lehum:2007nf,Lehum:2010tt,Queiruga:2015fzn},
while a study of a possible supersymmetry breaking would involve also
the knowledge of ${\cal F}$\,\cite{Gallegos:2011ux,Lehum:2013rpa}.

An explicit perturbative evaluation of $V_{eff}^{S}$ starting from
Eq.\,\eqref{eq:Lagran-Total-2}, within the superfield formalism,
is in general quite difficult. The root of this problem is the fact
that the classical superfield $\sigma_{cl}$ is spacetime constant,
but its covariant derivatives do not vanish, $D_{\alpha}\sigma\neq0$.
This, for example, complicates the calculation of the free superpropagators
of the model, since powers of $\sigma_{cl}$ appear in the quadratic
operators which have to be inverted. This leads to the appearance
of noncovariant superpropagators, as shown in\,\cite{Gallegos:2011ag,Gallegos:2011ux}.
In these works, the effective superpotential of the supersymmetric
Chern-Simons-matter and QED models was calculated up to two loops.
Part of these calculations was performed in an arbitrary gauge, but
the final results for the effective superpotential are obtained in
the Landau gauge. Another perspective on the difficulties of evaluating
the full effective superpotential in the superfield formalism, using
heat kernel techniques, can be found in\,\cite{Ferrari:2009zx}. 

One possibility to obtain a full computation of $V_{eff}^{S}$ within
the superfield formalism would involve the use of the Renormalization
Group Equation (RGE). This approach was used to calculate $K_{eff}$
in the massless limit of the model, in which case the action\,\eqref{eq:actionCSM}
is scale invariant, in\,\cite{Quinto2016}. Essentially, one may
consider $K_{eff}$ as a function of the single mass scale $\sigma_{1}$,
and imposes the RGE,
\begin{align}
\left[\mu\frac{\partial}{\partial\mu}+\beta_{x}\frac{\partial}{\partial x}-\gamma_{\Phi}\sigma_{1}\frac{\partial}{\partial\sigma_{1}}\right]K_{eff}\left(\sigma_{1};\mu,x,L\right) & =0\,,\label{eq:RGE1}
\end{align}
where $x$ generically denotes the coupling constants of the theory,
$\mu$ is the mass scale introduced by the regularization, 
\begin{align}
L & =\mbox{ln}\left[\frac{\sigma_{1}^{2}}{\mu}\right]\thinspace,\label{eq:defL}
\end{align}
and $\gamma_{\Phi}$ is the anomalous dimension of scalar superfield.
The scale invariance of the model constrains the form of the radiative
corrections to $K_{eff}$, so that a simple ansatz can be made, which
inserted in Eq.\,\eqref{eq:RGE1} provides a set of recursive equations
from which coefficients of the so-called leading logs contributions
to $K_{eff}$ can be found. This technique could be in principle extended
for the full effective superpotential $V_{eff}^{S}$, but in this
case more complicated, multiscale techniques would be needed\,\cite{Ford:1994dt,Ford:1996hd},
since $V_{eff}^{S}$ should be considered as a function both of $\sigma_{1}$
and $\sigma_{2}$. 

The derivation of the effective superpotential from the renormalization
group functions, by means of the RGE, may allow one to infer from
the gauge (in)dependence of the beta functions and anomalous dimensions
the gauge (in)dependence of the effective superpotential itself. This
is something we can do for the $K_{eff}$, since it was already established
in\,\cite{Quinto2016} how it can be calculated from the renormalization
group functions. Therefore, in the next section, we will present a
detailed computation of the beta and gamma functions in an arbitrary
gauge, showing that the result is indeed gauge independent. By this
reasoning, we can conclude that $K_{eff}$ do not depend on the gauge
parameter. That means, when only $K_{eff}$ is considered, the Nielsen
identity\,\eqref{eq:Identidade-Nielsen-V} is trivially satisfied
with $C^{S}=0$. 

These results may suggest the gauge independence of the whole effective
superpotential $V_{eff}^{S}$, but without explicitly establishing
that the RGE fixes the form of $\mathcal{F}$ in some approximation,
without ambiguities, from the renormalization group functions, which
we know are gauge independent, we believe this is still an open question.
The discussion of\,\cite{Gallegos:2011ag,Gallegos:2011ux} is not
conclusive in this regard, since most of the results are presented
in a specific gauge, but some gauge dependence was found in the effective
superpotential of the supersymmetric QED model. It is also not simple
to use the Nielsen identity itself to investigate this point since,
as discussed in\,\cite{Nielsen:1975fs,Aitchison:1983ns}, the calculation
of $C^{S}$ in the massless case is complicated by the fact that different
loop orders contribute to $C^{S}$ in a given order in the coupling
constants. 

\section{\label{sec:Calculation-of-beta-function}Gauge invariance of the
Renormalization Group Functions}

In this section, to establish the gauge independence of the renormalization
group functions of our model, we consider the $m=0$ version of Eq.\,\eqref{eq:actionCSM},
generalized to exhibit a global $SU\left(N\right)$ symmetry, 
\begin{eqnarray}
\mathcal{S}_{CS} & = & \int d^{5}z\left\{ -\frac{1}{2}\Gamma^{\alpha}W_{\alpha}-\frac{1}{2}\overline{\nabla^{\alpha}\Phi_{a}}\nabla_{\alpha}\Phi_{a}+\frac{\lambda}{4}\left(\overline{\Phi_{a}}\Phi_{a}\right)^{2}\right\} ,\label{eq:M1}
\end{eqnarray}
where the $N$ matter superfields carry indices of the fundamental
representation of the $SU\left(N\right)$ group. We introduce the
gauge-fixing action,
\begin{align}
\mathcal{S}_{GF} & =\frac{1}{4\,\alpha}\int d^{5}z\left(D^{\alpha}\Gamma_{\alpha}\right)^{2}\thinspace,\label{eq:GF}
\end{align}
but differently from what is done in the literature\,\cite{Avdeev:1991za,lehum:2007nf,Lehum:2010tt},
we will perform all calculations with an arbitrary gauge-fixing parameter.
From Eqs.\,\eqref{eq:M1} and\,\eqref{eq:GF} we have,
\begin{align}
\mathcal{S} & =\int d^{5}z\left\{ \frac{1}{4}\Gamma_{\alpha}\left[D^{\beta}D^{\alpha}+\frac{1}{\alpha}\,D^{\alpha}D^{\beta}\right]\Gamma_{\beta}-\frac{1}{2}\,g^{2}C_{\beta\alpha}\Gamma^{\alpha}\Gamma^{\beta}\overline{\Phi}_{a}\Phi_{a}\right.\nonumber \\
 & \left.+\overline{\Phi}_{a}D^{2}\Phi_{a}-\frac{i}{2}\,g\left[\Gamma^{\alpha}\overline{\Phi}_{a}D_{\alpha}\Phi_{a}-\left(D^{\alpha}\overline{\Phi}_{a}\right)\Gamma_{\alpha}\Phi_{a}\right]+\frac{\lambda}{4}\left(\overline{\Phi}_{a}\Phi_{a}\right)^{2}+\mathcal{L}_{ct}\right\} ,\label{eq:M3}
\end{align}
where the counterterm Lagrangian is,
\begin{align}
\mathcal{L}_{ct} & =\frac{\left(Z_{\Gamma}-1\right)}{4}\Gamma_{\alpha}D^{\beta}D^{\alpha}\Gamma_{\beta}+\frac{\left(Z_{\Phi}-1\right)}{2}\overline{\nabla^{\alpha}\Phi_{a}}\nabla_{\alpha}\Phi_{a}+\frac{\lambda}{4}\,Z_{\lambda}\left(\overline{\Phi_{a}}\Phi_{a}\right)^{2},\label{eq:M3CT}
\end{align}
$Z_{\Gamma}$, $Z_{\Phi}$ and $Z_{\lambda}$ being the counterterms
needed to make the renormalized quantities finite in each order of
perturbation theory. From Eq.\,\eqref{eq:M3} it follows the scalar
and gauge propagators,
\begin{equation}
\left\langle \overline{\Phi}_{i}\left(k,\theta_{1}\right)\Phi_{j}\left(-k,\theta_{2}\right)\right\rangle =i\delta_{ij}\frac{D^{2}}{k^{2}}\delta^{2}\left(\theta_{1}-\theta_{2}\right),\label{eq:ProgatorScalar}
\end{equation}
and
\begin{equation}
\left\langle \Gamma_{\beta}\left(k,\theta_{1}\right)\Gamma_{\rho}\left(-k,\theta_{2}\right)\right\rangle =\frac{i}{2k^{2}}\left(D_{\rho}D_{\beta}+\alpha\,D_{\beta}D_{\rho}\right)\delta^{2}\left(\theta_{1}-\theta_{2}\right)\,,\label{eq:PropagatorGauge}
\end{equation}
where 
\begin{equation}
D_{\rho}D_{\beta}+\alpha\,D_{\beta}D_{\rho}=b\left(\alpha\right)\,k_{\rho\beta}+a\left(\alpha\right)\,C_{\beta\alpha}\,D^{2}\thinspace,
\end{equation}
and 
\begin{equation}
a\left(\alpha\right)=1-\alpha\thinspace;\thinspace b\left(\alpha\right)\equiv1+\alpha\thinspace.\label{eq:defAB}
\end{equation}

\begin{figure}
\begin{centering}
\includegraphics[scale=0.6]{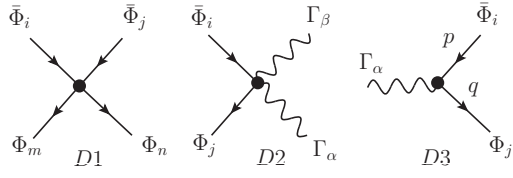}
\par\end{centering}
\caption{\label{fig:The-vertices-Diagrams}Elementary vertices of the model,
where a continuous line is associated to a scalar superfield, and
a wavy line to the gauge superfield. }
\end{figure}

The elementary vertices are represented in Figure\,\ref{fig:The-vertices-Diagrams}
and their analytic expression are
\begin{align}
i\,V_{\left(\overline{\Phi}_{a}\Phi_{a}\right)^{2}} & =2\,i\,\lambda\left(\delta_{in}\delta_{jm}+\delta_{im}\delta_{jn}\right),\label{eq:Vertex4PointsScalar}
\end{align}

\begin{align}
i\,V_{\overline{\Phi}_{a}\Phi_{a}\Gamma_{\beta}\Gamma_{\alpha}} & =i\,g^{2}\,\delta_{ij}\,C^{\alpha\beta},\label{eq:Vertex4PointGauge-Scalar}
\end{align}
and
\begin{align}
i\,V_{\Phi_{a}D^{\alpha}\overline{\Phi}_{a}\Gamma_{\alpha}-\Phi_{a}D^{\alpha}\Phi_{a}\Gamma_{\alpha}} & =\frac{1}{2}\,\delta_{ij}\,g\,\left[D^{\alpha}\left(p\right)-D^{\alpha}\left(-q\right)\right],\label{eq:Vertex3PointGauge-Scalar}
\end{align}
where an integration in the Grassmann coordinate of the superspace
is omitted at each vertex.

The use of regularization by dimensional reduction means that all
super-algebra manipulations are performed in three dimensions, while
momentum integrals are calculated at dimension $d=3-\epsilon$. The
use of this regularization scheme guarantees that the one loop correction
are finite. All algebraic manipulations of supercovariant derivatives
needed for the evaluation of supergraphs were performed with the Mathematica
package SusyMath\,\cite{ferrarisusymath}; explicit details about
the calculations will be presented elsewhere\,\cite{Quintophdthesis}.
\begin{center}
\begin{figure}
\centering{}\includegraphics[scale=0.8]{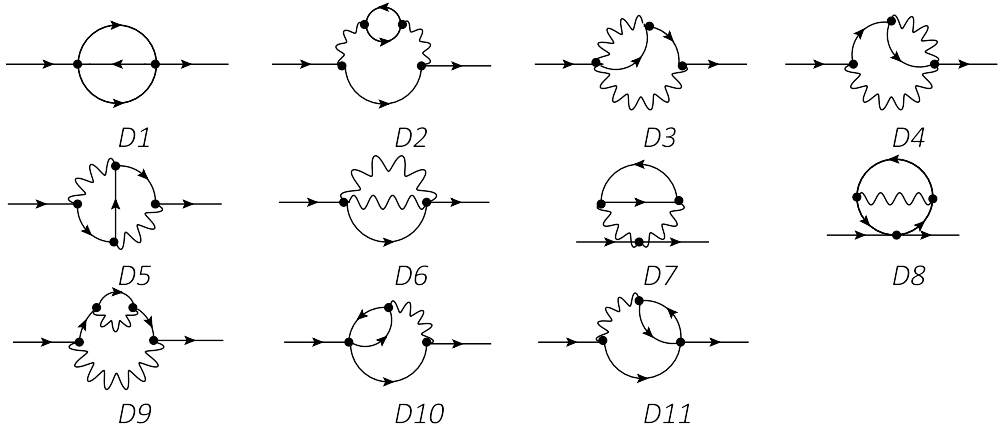}\caption{\label{fig:The-contribution-to-Two-Points-scalar}Two loops diagrams
contributing to the two-point vertex function of the scalar superfield
$\Phi$. }
\end{figure}
\par\end{center}

\begin{center}
\begin{table}
\centering{}%
\begin{tabular}{lccclccclccclcc}
 &  &  &  &  &  &  &  &  &  &  &  &  &  & \tabularnewline
\hline 
\hline 
$D1$ &  & $-2\left(N+1\right)\lambda^{2}$ &  & $D2$ &  & $\left(a+b\right)^{2}g^{4}N$ &  & $D3$ &  & $a\left(b-3\,a\right)g^{4}$ &  & $D4$ &  & $a\left(b-3\,a\right)\,g^{4}$\tabularnewline
$D5$ &  & $\frac{1}{2}\left(3\,a^{2}+2\,a\,b+3\,b^{2}\right)g^{4}$ &  & $D6$ &  & $2\,a^{2}\,g^{4}$ &  & $D7$ &  & $0$ &  & $D8$ &  & $0$\tabularnewline
$D9$ &  & $4\,a^{2}\,g^{4}$ &  & $D10$ &  & $0$ &  & $D11$ &  & $0$ &  &  &  & \tabularnewline
\hline 
\hline 
 &  &  &  &  &  &  &  &  &  &  &  &  &  & \tabularnewline
\end{tabular}\caption{\label{tab:Results-of-PhiPhi Diagrams}Divergent contributions from
each diagram presented in Figure\,\ref{fig:The-contribution-to-Two-Points-scalar},
with the common factor $\frac{1}{8}\left(\frac{i}{32\pi^{2}\epsilon}\right)\int\frac{d^{3}p}{\left(2\pi\right)^{3}}\,d^{2}\theta\,\overline{\Phi}_{i}\left(p,\theta\right)D^{2}\Phi_{i}\left(-p,\theta\right)$
omitted.}
\end{table}
\par\end{center}

We start by calculating the two-point vertex functions associated
to scalar and gauge superfields. For the case of the scalar superfield,
the diagrams that contribute are represented in the Figure\,\ref{fig:The-contribution-to-Two-Points-scalar},
and the corresponding results are given in Table\,\ref{tab:Results-of-PhiPhi Diagrams}.
For the divergent part of the two points vertex function, up to two
loops, we can write
\begin{align}
S_{\bar{\Phi}\Phi}^{2\,loop} & =\frac{i}{4\left(32\pi^{2}\epsilon\right)}\left[-\left(N+1\right)\lambda^{2}+\frac{1}{4}\left(a+b\right)^{2}\left(2\,N+3\right)g^{4}\right]\int\frac{d^{3}p}{\left(2\pi\right)^{3}}\,d^{2}\theta\,\overline{\Phi}_{i}\left(p,\theta\right)D^{2}\Phi_{i}\left(-p,\theta\right),\label{eq:PhiPhi contribution}
\end{align}
which fixes the value of the $Z_{\Phi}$ counterterm as
\begin{align}
Z_{\Phi} & =1+\frac{i}{4\left(32\pi^{2}\epsilon\right)}\left[-\left(N+1\right)\lambda^{2}+\frac{1}{4}\left(a+b\right)^{2}\left(2\,N+3\right)g^{4}\right].\label{eq:Ct ZPhi}
\end{align}
Remembering Eq.\,\eqref{eq:defAB}, we see that $Z_{\Phi}$ depends
on the gauge independent combination $a+b$. From this, it follows
that the anomalous dimension $\gamma_{\Phi}$ will also be gauge independent.
\begin{center}
\begin{figure}
\begin{centering}
\includegraphics[scale=0.7]{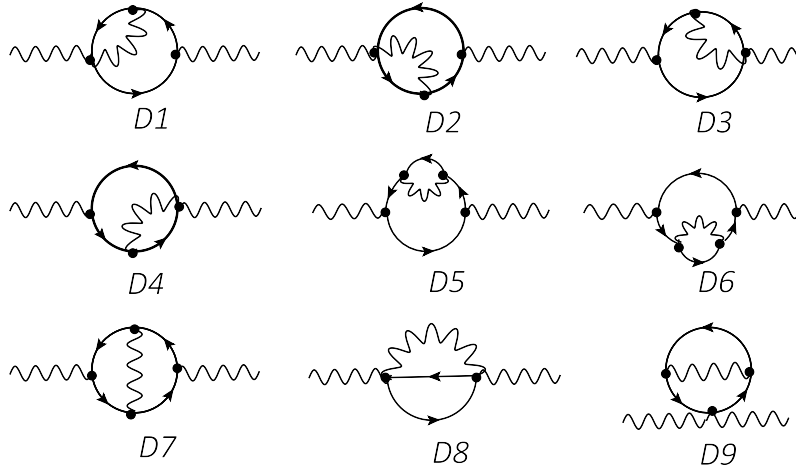}
\par\end{centering}
\caption{\label{fig:The-contribution-to-Two-Points-gauge}Two loop diagrams
contributing to the two-point vertex function of the gauge superfield
$\Gamma_{\alpha}$. }
\end{figure}
\par\end{center}

\begin{center}
\begin{table}
\centering{}%
\begin{tabular}{lcccccc}
 &  &  &  &  &  & \tabularnewline
\hline 
\hline 
$D1$ &  & $\left(3\,a-b\right)\left\{ -p_{\alpha\beta}\,+3\,C_{\beta\alpha}\,D^{2}\right\} $ &  & $D2$ &  & $\left(3\,a-b\right)\left\{ -p_{\alpha\beta}\,+3\,C_{\beta\alpha}\,D^{2}\right\} $\tabularnewline
$D3$ &  & $\left(3\,a-b\right)\left\{ -p_{\alpha\beta}\,+3\,C_{\beta\alpha}\,D^{2}\right\} $ &  & $D4$ &  & $\left(3\,a-b\right)\left\{ -p_{\alpha\beta}\,+3\,C_{\beta\alpha}\,D^{2}\right\} $\tabularnewline
$D5$ &  & $-2\,a\left\{ p_{\alpha\beta}\,+3\,C_{\beta\alpha}\,D^{2}\right\} $ &  & $D6$ &  & $-2\,a\left\{ p_{\alpha\beta}\,+3\,C_{\beta\alpha}\,D^{2}\right\} $\tabularnewline
$D7$ &  & $4\left\{ \left(4\,a+b\right)p_{\alpha\beta}+3\,b\,C_{\beta\alpha}D^{2}\right\} $ &  & $D8$ &  & $-\left\{ b\,p_{\alpha\beta}+3\,a\,C_{\beta\alpha}D^{2}\right\} $\tabularnewline
$D9$ &  & $0$ &  &  &  & \tabularnewline
\hline 
\hline 
 &  &  &  &  &  & \tabularnewline
\end{tabular}\caption{\label{tab:Result-of-GammaGamma}Divergent contributions from each
diagram in Figure\,\ref{fig:The-contribution-to-Two-Points-gauge},
omitting the common factor $\frac{1}{8}\left(\frac{N}{192\pi^{2}\epsilon}\right)i\,g^{4}\int\frac{d^{3}p}{\left(2\pi\right)^{3}}\,d^{2}\theta\,\Gamma^{\alpha}\left(p,\theta\right)\Gamma^{\beta}\left(-p,\theta\right)$.}
\end{table}
\par\end{center}

The next step is to compute up to two loops the two-point vertex function
of the gauge superfield $\Gamma_{\alpha}$. The diagrams involved
are represented in Figure\,\ref{fig:The-contribution-to-Two-Points-gauge},
with the respective divergent contributions given in Table\,\ref{tab:Result-of-GammaGamma}.
We verify that, for any gauge choice, all divergences cancel among
the graphs in Fig.\,\ref{fig:The-contribution-to-Two-Points-gauge}.
As a consequence, no infinite wave function renormalization of the
CS superfield is needed, and therefore the anomalous dimension $\gamma_{\Gamma}$
vanishes. This result extends for the massless matter case according
to the Coleman-Hill theorem\,\cite{Coleman:1985zi}, and it was also
verified in previous calculations performed in a specific gauge\,\cite{Avdeev:1991za},
as well as in the non-supersymmetric version of the model\,\cite{dias:2003pw}.
\begin{center}
\begin{figure}
\centering{}\includegraphics[scale=0.8]{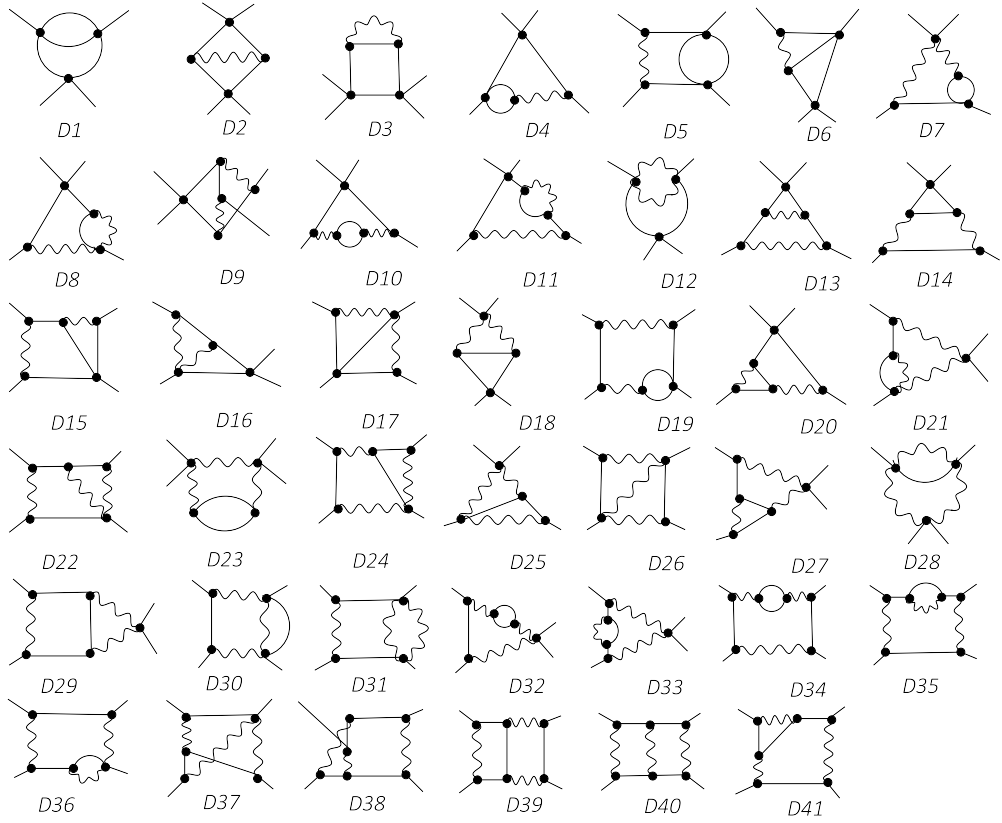}\caption{\label{fig:The-contribution-to-Four-Points-scalar}The complete set
of two loops diagrams contributing to the four-point scalar vertex
function.}
\end{figure}
\par\end{center}

\begin{center}
\begin{table}
\centering{}%
\begin{tabular}{lccclccclcc}
 &  &  &  &  &  &  &  &  &  & \tabularnewline
\hline 
\hline 
$D1$ &  & $-\frac{1}{4}\left(5N+11\right)\lambda^{3}$ &  & $D2$ &  & $\frac{1}{4}\,a\left(N+2\right)\lambda^{2}\,g^{2}$ &  & $D3$ &  & $-\frac{1}{4}\,a\left(N+4\right)\lambda^{2}\,g^{2}$\tabularnewline
$D4$ &  & $0$ &  & $D5$ &  & $-\frac{1}{4}\left(a-b\right)\lambda^{2}\,g^{2}$ &  & $D6$ &  & $\frac{1}{2}\left(a-b\right)\lambda^{2}\,g^{2}$\tabularnewline
$D7$ &  & $0$ &  & $D8$ &  & $\frac{1}{8}\left(b^{2}-4\,a\,b+3\,a^{2}\right)\lambda\,g^{4}$ &  & $D9$ &  & $\frac{3}{16}\left(a-b\right)^{2}\lambda\,g^{4}$\tabularnewline
$D10$ &  & $0$ &  & $D11$ &  & $\frac{1}{4}\,a\left(b-a\right)\lambda\,g^{4}$ &  & $D12$ &  & $\frac{3}{4}\,a^{2}\lambda\,g^{4}$\tabularnewline
$D13$ &  & $\frac{3}{4}\,a\,\left(a-b\right)\lambda\,g^{4}$ &  & $D14$ &  & $0$ &  & $D15$ &  & $-\frac{1}{4}\left(a-b\right)^{2}\lambda\,g^{4}$\tabularnewline
$D16$ &  & $\frac{3}{2}\,a\,\left(b-a\right)\lambda\,g^{4}$ &  & $D16$ &  & $\frac{3}{2}\,a\,\left(b-a\right)\lambda\,g^{4}$ &  & $D17$ &  & $\frac{1}{4}\left(a-b\right)^{2}\lambda\,g^{4}$\tabularnewline
$D18$ &  & $\frac{1}{8}\left(a+b\right)^{2}\left(1+N\right)\lambda\,g^{4}$ &  & $D19$ &  & $0$ &  & $D20$ &  & $-\frac{1}{8}\left(a-b\right)^{2}\lambda\,g^{4}$\tabularnewline
$D21$ &  & $-\frac{1}{8}\left(a-b\right)^{2}\left(3\,a-b\right)g^{6}$ &  & $D22$ &  & $0$ &  & $D23$ &  & $\frac{1}{16}\left(a+b\right)^{3}N\,g^{6}$\tabularnewline
$D24$ &  & $\frac{1}{8}\left(a-b\right)^{3}g^{6}$ &  & $D25$ &  & $-\frac{1}{4}\left(a-b\right)\left(a^{2}+b^{2}\right)g^{6}$ &  & $D26$ &  & $0$\tabularnewline
$D27$ &  & $\frac{1}{8}\left(a-b\right)^{3}g^{6}$ &  & $D28$ &  & $\frac{1}{2}\,a\,\left(a^{2}+2\,b^{2}\right)g^{6}$ &  & $D29$ &  & $\frac{1}{16}\left(a-b\right)\left(a+b\right)^{2}g^{6}$\tabularnewline
$D30$ &  & $-\frac{1}{8}\left(a-b\right)^{2}\left(2\,a-b\right)g^{6}$ &  & $D31$ &  & $0$ &  & $D32$ &  & $0$\tabularnewline
$D33$ &  & $\frac{1}{8}\,a\left(a-b\right)^{2}g^{6}$ &  & $D34$ &  & $0$ &  & $D35$ &  & $-\frac{1}{8}\,a\left(a-b\right)^{2}g^{6}$\tabularnewline
$D36$ &  & $\frac{1}{8}\left(a-b\right)^{2}\left(3\,a-b\right)g^{6}$ &  & $D37$ &  & $0$ &  & $D38$ &  & $0$\tabularnewline
$D39$ &  & $0$ &  & $D40$ &  & $0$ &  & $D41$ &  & $\frac{1}{8}\left(b-a\right)^{3}g^{6}$\tabularnewline
\hline 
\hline 
 &  &  &  &  &  &  &  &  &  & \tabularnewline
\end{tabular}\caption{\label{tab:Result-of-Four-Point}Divergent contributions arising from
the diagrams presented in Figure\,\ref{fig:The-contribution-to-Four-Points-scalar};
all contributions include the common factor $\frac{i}{32\pi^{2}\epsilon}\left\{ \delta_{im}\,\delta_{nj}+\delta_{jm}\,\delta_{ni}\right\} \int d^{2}\theta\,\Phi_{n}\left(0,\theta\right)\Phi_{m}\left(0,\theta\right)\overline{\Phi}_{i}\left(0,\theta\right)\overline{\Phi}_{j}\left(0,\theta\right)$,
where all external momenta were set to zero.}
\end{table}
\par\end{center}

Finally, the evaluation of the divergent part of the four-point vertex
function associated to the scalar superfield $\Phi$, up to two loops,
involve all diagrams in the Figure\,\ref{fig:The-contribution-to-Four-Points-scalar}.
The results are given in Table\,\ref{tab:Result-of-Four-Point},
and lead to
\begin{align}
S_{\left(\bar{\Phi}\Phi\right)^{2}}^{2\,loop} & =\frac{i}{4\left(32\pi^{2}\epsilon\right)}\left[-\left(5\,N+11\right)\lambda^{3}-\left(a+b\right)\lambda^{2}\,g^{2}+\frac{1}{4}\left(a+b\right)^{2}\left(2\,N+5\right)\lambda\,g^{4}\right.\nonumber \\
 & \left.+\frac{1}{4}\left(a+b\right)^{3}\left(N+3\right)g^{6}\right]\left\{ \delta_{im}\,\delta_{nj}+\delta_{jm}\,\delta_{ni}\right\} \int d^{2}\theta\,\Phi_{n}\left(0,\theta\right)\Phi_{m}\left(0,\theta\right)\overline{\Phi}_{i}\left(0,\theta\right)\overline{\Phi}_{j}\left(0,\theta\right)\thinspace,\label{eq:PhiPhiPhiPhi contribution}
\end{align}
which implies that
\begin{align}
Z_{\lambda} & =1+\frac{1}{2\left(32\pi^{2}\epsilon\right)}\left[\left(5\,N+11\right)\lambda^{2}+\left(a+b\right)\lambda\,g^{2}-\frac{1}{4}\left(a+b\right)^{2}\left(2\,N+5\right)g^{4}\right.\nonumber \\
 & \left.-\frac{1}{4}\left(a+b\right)^{3}\left(N+3\right)\lambda^{-1}g^{6}\right].\label{eq:Ct ZLambda}
\end{align}

In conclusion, all the renormalization constants $Z_{\Phi}$, $Z_{\Gamma}$
and $Z_{\lambda}$ are independent of the choice of the gauge-fixing
parameter. This same property will follow for the renormalization
group functions that are calculated from these constants, and from
the procedure described in\,\cite{Quinto2016}, to the effective
superpotential $K_{eff}$.

The explicit relations between bare and renormalized quantities are
given by
\begin{align}
\Gamma_{0}^{\alpha} & =Z_{\Gamma}^{\frac{1}{2}}\Gamma^{\alpha},\label{eq:ZGamma}\\
\Phi_{0} & =Z_{\Phi}^{\frac{1}{2}}\Phi,\label{eq:ZPhi}\\
\alpha_{0} & =\alpha\,Z_{\Gamma},\label{eq:Zalpha}\\
g_{0} & =\mu^{\frac{\epsilon}{2}}g\,Z_{\Gamma}^{-\frac{1}{2}},\label{eq:Zg}\\
\lambda_{0} & =\mu^{\epsilon}\lambda\,Z_{\lambda}Z_{\Phi}^{-2},\label{eq:Zlambda}
\end{align}
where $\mu$ is a mass parameter introduced to keep $g$ and $\lambda$
dimensionless. From these, follow the renormalization group functions,
\begin{align}
\gamma_{\Gamma} & \equiv-\frac{\mu}{\Gamma}\frac{d}{d\mu}\Gamma=\frac{\mu}{2\,Z_{\Gamma}}\frac{d}{d\mu}Z_{\Gamma},\label{eq:gammadeGamma}\\
\gamma_{\Phi} & =\frac{\mu}{2\,Z_{\Phi}}\frac{d}{d\mu}Z_{\Phi},\label{eq:gammadePhi}\\
\beta_{\alpha} & \equiv\mu\frac{d}{d\mu}\alpha=-2\,\alpha\,\gamma_{\Gamma},\label{eq:BetaAlpha}\\
\beta_{g} & =g\,\gamma_{\Gamma},\label{eq:Beta g}\\
\beta_{\lambda} & =\epsilon\frac{\lambda^{2}}{Z_{\lambda}}\frac{\partial}{\partial\lambda}Z_{\lambda}+\frac{\epsilon}{2}\frac{\lambda\,g}{Z_{g}}\frac{\partial}{\partial g}Z_{g}+4\lambda\gamma_{\Phi}.\label{eq:Beta lambda}
\end{align}
From these definitions, and the results presented in this section,
we obtain\textbf{
\begin{align}
\gamma_{\Phi} & =\frac{1}{4\left(32\pi^{2}\right)}\left\{ \left(N+1\right)\lambda^{2}-\left(2\,N+3\right)g^{4}\right\} ,\label{eq:gamma de Phi Final}
\end{align}
}
\begin{align}
\gamma_{\Gamma}=\beta_{\alpha}=\beta_{g} & =0,\label{eq:gamma de Gamma Final}
\end{align}
and
\begin{align}
\beta_{\lambda} & =\frac{1}{16\pi^{2}}\left\{ 3\left(N+2\right)\lambda^{3}+\lambda^{2}\,g^{2}-2\left(N+2\right)\lambda\,g^{4}-\left(N+3\right)g^{6}\right\} .\label{eq:Beta Lambda final}
\end{align}

As stated before, these functions are independent of the gauge-fixing
parameter. Our results agree with reference\,\cite{Avdeev:1991za},
except for overall numerical factors that arise due to their different
definition of the scale $\mu$. We stress however that in\,\cite{Avdeev:1991za}
the authors considered as true the gauge independence of the these
functions, a fact that has to be checked explicitly.

\section{\label{sec:Conclusion}Conclusion}

The gauge independence of physical observables is an essential point
to consider when studying gauge theories. In relation to the effective
potential, which plays an essential role in the mechanism of symmetry
breaking in many relevant models, the Nielsen identity is the key
to understand how the effective potential can depend on the gauge
choice, and yet physical quantities evaluated at its minima can be
gauge independent.

In this paper we studied the Nielsen identity for a supersymmetric
Chern-Simons model in the superfield formalism. After deriving the
Nielsen identity in the superfield language, we argue that an explicit
calculation of the complete effective superpotential $V_{eff}^{S}$,
including any possible gauge dependence, is still a technically difficult
task. As a first step in this direction, we consider the part of the
effective superpotential which do not depend on supercovariant derivatives
of the background scalar superfield, $K_{eff}$. It was already shown
in\,\cite{Quinto2016} that $K_{eff}$ can be calculated from the
renormalization group functions, using the RGE. We then verify, by
means of an explicit calculation, that the beta and gamma functions
are gauge independent and, therefore, so is $K_{eff}$. Our results
agree with the ones previously found in the literature, that were
calculated in a specific gauge. 

As a future perspective, we will try to extend these results to the
full effective superpotential $V_{eff}^{S}$. This should be possible
with the results given in this article, and same techniques used in\,\cite{Quinto2016},
but generalized to the multiscale case\,\cite{Ford:1994dt,Ford:1996hd}.
If these techniques are shown to be robust enough to prove that all
terms in $V_{eff}^{S}$ are fixed in terms of the coefficients of
the beta and gamma functions, we may be finally able to establish
the gauge independence of the full effective superpotential $V_{eff}^{S}$.

Also, a generalization of this study for non Abelian models would
be an interesting endeavor. Already at the perturbative level, in
a non Abelian model the computation of the renormalization group functions
would involve several additional diagrams, and even the ones already
existent will be modified by group theoretical factors. So, it is
not obvious that the dependence on the gauge parameter $\alpha$,
that cancelled among all diagrams in the Abelian case, as we shown,
will also cancel in the non Abelian case. This would make the study
of the Nielsen Identity much richer, since this identity would then
be essential do prove the gauge invariance of the physical properties
of the symmetry breaking. Finally, in non Abelian gauge theories there
are also non perturbative aspects that are relevant, e.g., Gribov
copies. These are profound questions that deserves more study.

\bigskip{}

\textbf{Acknowledgments. }This work was supported by Conselho Nacional
de Desenvolvimento Científico e Tecnológico (CNPq), Fundação de Amparo
a Pesquisa do Estado de São Paulo (FAPESP) and Coordenação de Aperfeiçoamento
de Pessoal de Nível Superior (CAPES), via the following grants: CNPq
482874/2013-9, FAPESP 2013/22079-8 and 2014/24672-0 (AFF), CAPES PhD
grant (AGQ).

\appendix

\section*{\label{sec:Two-Loops-Integrals}Two Loops Integrals}

In obtaining the results in this paper, we have used the following
two loops integrals in Minkowski spacetime, with spacetime metric
$\eta^{\alpha\beta}\equiv\mbox{diag}\left(-,+,+\right)$, where $d^{D}l\equiv\mu^{\epsilon}d^{3-\epsilon}l$,
$D=3-\epsilon$, and $C_{\alpha\beta}$ is the antisymmetric tensor
used to lower and raise spinor indices\,\cite{gates:1983nr}.
\begin{align}
\mathcal{I}_{1} & =\int\frac{d^{D}kd^{D}q}{\left(2\pi\right)^{2D}}\frac{1}{\left(k+p\right)^{2}\left(q+k\right)^{2}q^{2}}=-\frac{1}{32\pi^{2}\epsilon},\label{eq:Int 1}
\end{align}
\begin{align}
\mathcal{I}_{2}=\int\frac{d^{D}kd^{D}q}{\left(2\pi\right)^{2D}}\frac{q_{\alpha\beta}}{\left(k+p\right)^{2}\left(q+k\right)^{2}q^{2}} & =-\frac{p_{\alpha\beta}}{96\pi^{2}\epsilon},\label{eq:Int 2}
\end{align}

\begin{align}
\mathcal{I}_{3} & =\int\frac{d^{D}kd^{D}q}{\left(2\pi\right)^{2D}}\frac{k_{\alpha\beta}}{\left(k+p\right)^{2}\left(q+k\right)^{2}q^{2}}=\frac{p_{\alpha\beta}}{48\pi^{2}\epsilon},\label{eq:Int 3}
\end{align}

\begin{align}
\mathcal{I}_{4} & =\int\frac{d^{D}kd^{D}q}{\left(2\pi\right)^{2D}}\frac{k_{\alpha\beta}\,q_{\theta\lambda}}{\left(k+p\right)^{2}k^{2}\left(q+k\right)^{2}q^{2}}=\frac{1}{192\pi^{2}\epsilon}\left(C_{\alpha\theta}\,C_{\beta\lambda}+C_{\beta\theta}\,C_{\alpha\lambda}\right)\label{eq:Int 4}
\end{align}
\begin{align}
\mathcal{I}_{5} & =\int\frac{d^{D}kd^{D}q}{\left(2\pi\right)^{2D}}\frac{k_{\mu\nu}\,k_{\theta\lambda}}{\left(k+p\right)^{2}k^{2}\left(k+q\right)^{2}q^{2}}=-\frac{1}{96\pi^{2}\epsilon}\left(C_{\mu\theta}\,C_{\nu\lambda}+C_{\nu\theta}\,C_{\mu\lambda}\right),\label{eq:Int 5}
\end{align}

\begin{align}
\mathcal{I}_{6} & =\int\frac{d^{D}kd^{D}q}{\left(2\pi\right)^{2D}}\frac{q_{\lambda\theta}\,k_{\beta\rho}}{\left(k+p\right)^{2}\left(p+q\right)^{2}\left(k-q\right)^{2}q^{2}}=\int\frac{d^{D}kd^{D}q}{\left(2\pi\right)^{2D}}\frac{q_{\lambda\theta}\,k_{\beta\rho}}{\left(k+p\right)^{2}\left(p+q\right)^{2}\left(k-q\right)^{2}k^{2}}\nonumber \\
 & =-\frac{1}{192\pi^{2}\epsilon}\left(C_{\lambda\beta}\,C_{\theta\rho}+C_{\theta\beta}\,C_{\lambda\rho}\right),\label{eq: Int 6}
\end{align}

\begin{align}
\mathcal{I}_{7} & =\int\frac{d^{D}kd^{D}q}{\left(2\pi\right)^{2D}}\frac{q_{\mu\beta}\,q_{\theta\lambda}}{\left(k+p\right)^{2}\left(p+q\right)^{2}\left(k-q\right)^{2}q^{2}}=\int\frac{d^{D}kd^{D}q}{\left(2\pi\right)^{2D}}\frac{k_{\mu\beta}\,k_{\theta\lambda}}{\left(k+p\right)^{2}\left(p+q\right)^{2}\left(k-q\right)^{2}k^{2}}\nonumber \\
 & =-\frac{1}{96\pi^{2}\epsilon}\left(C_{\mu\theta}\,C_{\beta\lambda}+C_{\beta\theta}\,C_{\mu\lambda}\right),\label{eq: Int 7}
\end{align}
\begin{align}
\mathcal{I}_{8} & =\int\frac{d^{D}kd^{D}q}{\left(2\pi\right)^{2D}}\frac{q_{\mu\nu}\,q_{\lambda\theta}\,k_{\rho\sigma}}{\left(k+p\right)^{2}\left(q+p\right)^{2}\left(k-q\right)^{2}q^{2}}=\int\frac{d^{D}kd^{D}q}{\left(2\pi\right)^{2D}}\frac{k_{\mu\nu}\,k_{\lambda\theta}\,q_{\rho\sigma}}{\left(k+p\right)^{2}\left(q+p\right)^{2}\left(k-q\right)^{2}k^{2}}\nonumber \\
 & =\frac{1}{320\pi^{2}\epsilon}\left\{ \left(C_{\lambda\rho}\,C_{\theta\sigma}+C_{\theta\rho}\,C_{\lambda\sigma}\right)p_{\mu\nu}+\left(C_{\mu\rho}\,C_{\nu\sigma}+C_{\nu\rho}\,C_{\mu\sigma}\right)p_{\lambda\theta}+\frac{8}{3}\left(C_{\mu\lambda}\,C_{\nu\theta}+C_{\nu\lambda}\,C_{\mu\theta}\right)p_{\rho\sigma}\right\} ,\label{eq: Int 8}
\end{align}

\begin{align}
\mathcal{I}_{9} & =\int\frac{d^{D}kd^{D}q}{\left(2\pi\right)^{2D}}\frac{q_{\mu\lambda}\,k_{\zeta\theta}\,k_{\kappa\rho}}{\left(k+p\right)^{2}k^{2}\left(k+q\right)^{2}q^{2}}=-\frac{1}{480\pi^{2}\epsilon}\left\{ \left(C_{\mu\zeta}\,C_{\lambda\theta}+C_{\lambda\zeta}\,C_{\mu\theta}\right)p_{\kappa\rho}\right.\nonumber \\
 & \left.+\left(C_{\zeta\kappa}\,C_{\theta\rho}+C_{\theta\kappa}\,C_{\zeta\rho}\right)p_{\mu\lambda}+\left(C_{\mu\kappa}\,C_{\lambda\rho}+C_{\lambda\kappa}\,C_{\mu\rho}\right)p_{\zeta\theta}\right\} ,\label{eq: Int 9}
\end{align}

\begin{align}
\mathcal{I}_{10} & =\int\frac{d^{D}kd^{D}q}{\left(2\pi\right)^{2D}}\frac{k_{\lambda\theta}\,k_{\mu\nu}\,k_{\kappa\rho}\,q_{\sigma\gamma}}{\left(k+p\right)^{2}k^{2}\left(k+q\right)^{2}k^{2}q^{2}}=\frac{1}{960\pi^{2}\epsilon}\left\{ \left(C_{\lambda\mu}\,C_{\theta\nu}+C_{\theta\mu}\,C_{\lambda\nu}\right)\left(C_{\kappa\sigma}\,C_{\rho\gamma}+C_{\rho\sigma}\,C_{\kappa\rho}\right)\right.\nonumber \\
 & \left.+\left(C_{\lambda\kappa}\,C_{\theta\rho}+C_{\theta\kappa}\,C_{\lambda\rho}\right)\left(C_{\mu\sigma}\,C_{\nu\gamma}+C_{\nu\sigma}\,C_{\mu\gamma}\right)+\left(C_{\lambda\sigma}\,C_{\theta\gamma}+C_{\theta\sigma}\,C_{\lambda\gamma}\right)\left(C_{\mu\kappa}\,C_{\nu\rho}+C_{\nu\kappa}\,C_{\mu\rho}\right)\right\} ,\label{eq: Int 10}
\end{align}

\begin{align}
\mathcal{I}_{11} & =\int\frac{d^{D}kd^{D}q}{\left(2\pi\right)^{2D}}\frac{k_{\delta\theta}\,k_{\sigma\rho}\,q_{\beta\gamma}\,q_{\alpha\lambda}}{\left(k+p\right)^{2}\left(p+q\right)^{2}\left(k-q\right)^{2}k^{2}q^{2}}=-\frac{1}{5\left(384\pi^{2}\epsilon\right)}\left\{ 6\left(C_{\delta\sigma}\,C_{\theta\rho}+C_{\theta\sigma}\,C_{\delta\rho}\right)\times\right.\nonumber \\
 & \left(C_{\beta\alpha}\,C_{\gamma\lambda}+C_{\gamma\alpha}\,C_{\beta\lambda}\right)+\left(C_{\delta\beta}\,C_{\theta\gamma}+C_{\theta\beta}\,C_{\delta\gamma}\right)\left(C_{\sigma\alpha}\,C_{\rho\lambda}+C_{\rho\alpha}\,C_{\sigma\lambda}\right)\nonumber \\
 & \left.+\left(C_{\delta\alpha}\,C_{\theta\lambda}+C_{\theta\alpha}\,C_{\delta\lambda}\right)\left(C_{\sigma\beta}\,C_{\rho\gamma}+C_{\rho\beta}\,C_{\sigma\gamma}\right)\right\} ,\label{eq: Int 11}
\end{align}

\begin{align}
\mathcal{I}_{12} & =\int\frac{d^{D}kd^{D}q}{\left(2\pi\right)^{2D}}\frac{k_{\delta\theta}\,k_{\sigma\rho}\,q_{\beta\gamma}\,q_{\alpha\lambda}}{\left(k+p\right)^{2}\left(q^{2}\right)^{2}\left[\left(k-q\right)^{2}\right]^{2}}=\int\frac{d^{D}kd^{D}q}{\left(2\pi\right)^{2D}}\frac{k_{\delta\theta}\,k_{\sigma\rho}\,q_{\beta\gamma}\,q_{\alpha\lambda}}{\left(k^{2}\right)^{2}\left(q+p\right)^{2}\left[\left(k-q\right)^{2}\right]^{2}}\nonumber \\
 & =\frac{1}{4\left(160\pi^{2}\epsilon\right)}\left\{ -\frac{2}{3}\left(C_{\delta\sigma}\,C_{\theta\rho}+C_{\theta\sigma}\,C_{\delta\rho}\right)\left(C_{\beta\alpha}\,C_{\gamma\lambda}+C_{\gamma\alpha}\,C_{\beta\lambda}\right)+\left(C_{\delta\beta}\,C_{\theta\gamma}+C_{\theta\beta}\,C_{\delta\gamma}\right)\times\right.\nonumber \\
 & \left.\left(C_{\sigma\alpha}\,C_{\rho\lambda}+C_{\rho\alpha}\,C_{\sigma\lambda}\right)+\left(C_{\delta\alpha}\,C_{\theta\lambda}+C_{\theta\alpha}\,C_{\delta\lambda}\right)\left(C_{\sigma\beta}\,C_{\rho\gamma}+C_{\rho\beta}\,C_{\sigma\gamma}\right)\right\} ,\label{eq: Int 12}
\end{align}


\begin{thebibliography}{10}

\bibitem{Jackiw:1974cv}
R.~Jackiw.
\newblock {Functional evaluation of the effective potential}.
\newblock {\em Phys.Rev.}, D9:1686, 1974.

\bibitem{Dolan1974}
L.~Dolan and R.~Jackiw.
\newblock Gauge-invariant signal for gauge-symmetry breaking.
\newblock {\em Phys. Rev. D}, 9:2904--2912, May 1974.

\bibitem{Frere1975}
J.~M. Frere and P.~Nicoletopoulos.
\newblock {Gauge Invariant Content of the Effective Potential}.
\newblock {\em Phys. Rev.}, D11:2332, 1975.

\bibitem{Nielsen:1975fs}
N.~K. Nielsen.
\newblock {On the Gauge Dependence of Spontaneous Symmetry Breaking in Gauge
  Theories}.
\newblock {\em Nucl. Phys.}, B101:173--188, 1975.

\bibitem{Fukuda:1975di}
Reijiro Fukuda and Taichiro Kugo.
\newblock {Gauge Invariance in the Effective Action and Potential}.
\newblock {\em Phys. Rev.}, D13:3469, 1976.

\bibitem{Aitchison:1983ns}
I.~J.~R. Aitchison and C.~M. Fraser.
\newblock {Gauge Invariance and the Effective Potential}.
\newblock {\em Annals Phys.}, 156:1, 1984.

\bibitem{Johnston1985}
D.~Johnston.
\newblock {Nielsen Identities in the 't Hooft Gauge}.
\newblock {\em Nucl. Phys.}, B253:687--700, 1985.

\bibitem{DoNascimento1987}
J.~R.~S. Do~Nascimento and D.~Bazeia.
\newblock {Gauge Invariance of the Effective Potential}.
\newblock {\em Phys. Rev.}, D35:2490--2494, 1987.

\bibitem{Breckenridge1995}
J.~C. Breckenridge, M.~J. Lavelle, and Thomas~G. Steele.
\newblock {The Nielsen identities for the two point functions of QED and QCD}.
\newblock {\em Z. Phys.}, C65:155--164, 1995.

\bibitem{Gambino2000}
Paolo Gambino and Pietro~Antonio Grassi.
\newblock {The Nielsen identities of the SM and the definition of mass}.
\newblock {\em Phys. Rev.}, D62:076002, 2000.

\bibitem{Iguri2001}
Sergio~M. Iguri and Francisco~D. Mazzitelli.
\newblock {Gauge fixing independence of test fields in Yang-Mills theories}.
\newblock 2001.

\bibitem{Gerhold2003}
A.~Gerhold and A.~Rebhan.
\newblock {Gauge dependence identities for color superconducting QCD}.
\newblock {\em Phys. Rev.}, D68:011502, 2003.

\bibitem{Lewandowski2013}
Adrian Lewandowski.
\newblock {Renormalization of Nielsen Identities}.
\newblock 2013.

\bibitem{Upadhyay2016}
Sudhaker Upadhyay.
\newblock {Ward and Nielsen Identities for ABJM Theory in ${\cal N}=1$
  Superspace}.
\newblock 2016.

\bibitem{Avdeev:1991za}
L.V. Avdeev, G.V. Grigorev, and D.I. Kazakov.
\newblock {Renormalizations in Abelian Chern-Simons field theories with
  matter}.
\newblock {\em Nucl.Phys.}, B382:561--580, 1992.

\bibitem{Avdeev:1992jt}
L.~V. Avdeev, D.~I. Kazakov, and I.~N. Kondrashuk.
\newblock {Renormalizations in supersymmetric and nonsupersymmetric nonAbelian
  Chern-Simons field theories with matter}.
\newblock {\em Nucl. Phys.}, B391:333--357, 1993.

\bibitem{Collins_Book}
John Collins.
\newblock {\em {Renormalization}}.
\newblock {Cambridge Monographs on Mathematical Physics}. Cambridge Univ. Pr.,
  1984.

\bibitem{Fazio2001}
A.~R. Fazio, V.~E.~R. Lemes, M.~S. Sarandy, and S.~P. Sorella.
\newblock {The Diagonal ghost equation Ward identity for Yang-Mills theories in
  the maximal Abelian gauge}.
\newblock {\em Phys. Rev.}, D64:085003, 2001.

\bibitem{Bell2013}
J.~M. Bell and J.~A. Gracey.
\newblock {Momentum subtraction scheme renormalization group functions in the
  maximal Abelian gauge}.
\newblock {\em Phys. Rev.}, D88(8):085027, 2013.

\bibitem{DiLuzio2014}
Luca Di~Luzio and Luminita Mihaila.
\newblock {On the gauge dependence of the Standard Model vacuum instability
  scale}.
\newblock {\em JHEP}, 06:079, 2014.

\bibitem{Bell2015}
J.~M. Bell and J.~A. Gracey.
\newblock {Maximal abelian and Curci-Ferrari gauges in momentum subtraction at
  three loops}.
\newblock {\em Phys. Rev.}, D92:125001, 2015.

\bibitem{lehum:2007nf}
A.~C. Lehum, A.~F. Ferrari, M.~Gomes, and A.~J. da~Silva.
\newblock {Spontaneous gauge symmetry breaking in a {SUSY} Chern-Simons model}.
\newblock {\em Phys.Rev.}, 76:105021, 2007.

\bibitem{Lehum2009}
A.~C. Lehum.
\newblock {Dynamical breaking of gauge symmetry in supersymmetric quantum
  electrodynamics in three-dimensional spacetime}.
\newblock {\em Phys. Rev.}, D79:025005, 2009.

\bibitem{Ferrari:2009zx}
A.F. Ferrari, M.~Gomes, A.C. Lehum, J.R. Nascimento, A.Yu. Petrov, et~al.
\newblock {On the superfield effective potential in three dimensions}.
\newblock {\em Phys.Lett.}, B678:500--503, 2009.

\bibitem{Ferrari:2010ex}
A.F. Ferrari, E.A. Gallegos, M.~Gomes, A.C. Lehum, J.R. Nascimento, et~al.
\newblock {Coleman-Weinberg mechanism in a three-dimensional supersymmetric
  Chern-Simons-Matter model}.
\newblock {\em Phys.Rev.}, D82:025002, 2010.

\bibitem{Queiruga:2015fzn}
J.~M. Queiruga, A.~C. Lehum, and Mir Faizal.
\newblock {Kahlerian effective potentials for Chern-Simons-matter theories}.
\newblock {\em Nucl. Phys.}, B902:58--68, 2016.

\bibitem{Quinto2016}
A.~G. Quinto, A.~F. Ferrari, and A.~C. Lehum.
\newblock {Renormalization group improvement and dynamical breaking of symmetry
  in a supersymmetric Chern-Simons-matter model}.
\newblock {\em Nucl. Phys.}, B907:664--677, 2016.

\bibitem{Lehum:2010tt}
A.C. Lehum and A.J. da~Silva.
\newblock {Spontaneous breaking of superconformal invariance in (2+1)D
  supersymmetric Chern-Simons-matter theories in the large N limit}.
\newblock {\em Phys.Lett.}, B693:393--398, 2010.

\bibitem{gates:1983nr}
S.~J. Gates, Marcus~T. Grisaru, M.~Rocek, and W.~Siegel.
\newblock {Superspace, or one thousand and one lessons in supersymmetry}.
\newblock {\em Front. Phys.}, 58:1--548, 1983.

\bibitem{buchbinder:1998qv}
I.~L. Buchbinder and S.~M. Kuzenko.
\newblock {\em {Ideas and methods of supersymmetry and supergravity: Or a walk
  through superspace}}.
\newblock Bristol, UK, 1998.
\newblock 656 p.

\bibitem{Coleman:1973jx}
Sidney~R. Coleman and Erick~J. Weinberg.
\newblock {Radiative Corrections as the Origin of Spontaneous Symmetry
  Breaking}.
\newblock {\em Phys.Rev.}, D7:1888--1910, 1973.

\bibitem{Kang1974}
J.~S. Kang.
\newblock {Gauge Invariance of the Scalar-Vector Mass Ratio in the
  Coleman-Weinberg Model}.
\newblock {\em Phys. Rev.}, D10:3455, 1974.

\bibitem{Becchi1974}
C.~Becchi, A.~Rouet, and R.~Stora.
\newblock {The Abelian Higgs-Kibble Model. Unitarity of the S Operator}.
\newblock {\em Phys. Lett.}, B52:344--346, 1974.

\bibitem{Gallegos:2011ux}
E.A. Gallegos and A.J. da~Silva.
\newblock {Dynamical (super)symmetry vacuum properties of the supersymmetric
  Chern-Simons-matter model}.
\newblock {\em Phys.Rev.}, D85:125012, 2012.

\bibitem{Lehum:2013rpa}
A.~C. Lehum and A.~J. da~Silva.
\newblock {Supersymmetry breaking in the three-dimensional nonlinear sigma
  model}.
\newblock {\em Phys. Rev.}, D88(6):067702, 2013.

\bibitem{Gallegos:2011ag}
E.A. Gallegos and A.J. da~Silva.
\newblock {Supergraph techniques for D=3,N=1 broken supersymmetric theories}.
\newblock {\em Phys.Rev.}, D84:065009, 2011.

\bibitem{Ford:1994dt}
Christopher Ford.
\newblock {Multiscale renormalization group improvement of the effective
  potential}.
\newblock {\em Phys. Rev.}, D50:7531--7537, 1994.

\bibitem{Ford:1996hd}
C.~Ford and C.~Wiesendanger.
\newblock {A Multiscale subtraction scheme and partial renormalization group
  equations in the O(N) symmetric phi**4 theory}.
\newblock {\em Phys. Rev.}, D55:2202--2217, 1997.

\bibitem{ferrarisusymath}
A.~F. Ferrari.
\newblock {SusyMath: {A} Mathematica package for quantum superfield
  calculations}.
\newblock {\em Comput. Phys. Commun.}, 176:334--346, 2007.

\bibitem{Quintophdthesis}
Andr\'es~G\'omez Quinto.
\newblock {\em Effective Superpotential and the Renormalization Group Equation
  in a Supersymmetric Chern-Simons-Matter Model in the Superfield Formalism}.
\newblock PhD thesis, Universidade Federal do ABC - UFABC, 2016.
\newblock To appear.

\bibitem{Coleman:1985zi}
Sidney~R. Coleman and Brian~Russell Hill.
\newblock {No More Corrections to the Topological Mass Term in QED in
  Three-Dimensions}.
\newblock {\em Phys. Lett.}, B159:184--188, 1985.

\bibitem{dias:2003pw}
Alex~G. Dias, M.~Gomes, and A.~J. da~Silva.
\newblock {Dynamical breakdown of symmetry in (2+1) dimensional model
  containing the Chern-Simons field}.
\newblock {\em Phys. Rev.}, D69:065011, 2004.

\end{thebibliography}
\end{document}